# The Importance of Education for Technological Development and the Role of Internet-Based Learning in Education

Ozdemir Cetin[1], Murat Cakiroglu[2], Cüneyt Bayılmış[3], Hüseyin Ekiz[4]

[1,2,4] Sakarya University, Faculty of Technical Education, Department of Electronics and Computer Education, 54187 Adapazarı

[3] Kocaeli University, Faculty of Technical Education, Department of Electronics and Computer Education, 41300 İzmit

[1]ocetin@sakarya.edu.tr   [2]muratc@sakarya.edu.tr   [3]bayilmis @kou.edu.tr   [4]ekiz@sakarya.edu.tr

**Abstract**

In today's world, many technologically advanced countries have realized that real power lies not in physical strength but in educated minds. As a result, every country has embarked on restructuring its education system to meet the demands of technology. As a country in the midst of these developments, we cannot remain indifferent to this transformation in education.

In the Information Age of the 21st century, rapid access to information is crucial for the development of individuals and societies. To take our place among the knowledge societies in a world moving rapidly towards globalization, we must closely follow technological innovations and meet the requirements of technology. This can be achieved by providing learning opportunities to anyone interested in acquiring education in their area of interest.

This study focuses on the advantages and disadvantages of internet-based learning compared to traditional teaching methods, the importance of computer usage in internet-based learning, negative factors affecting internet-based learning, and the necessary recommendations for addressing these issues. In today's world, it is impossible to talk about education without technology or technology without education.

## EDUCATION AND TEACHING

Education and teaching are two concepts that cannot be separated. Let's consider these two terms that are now constantly used interchangeably:

Education is the process of bringing about changes in individuals' lives and behaviors in line with predetermined goals. According to Ertürk (1975), "education is the process of intentionally and deliberately bringing about the desired change in an individual's behavior through their experiences."

Teaching, on the other hand, can be defined as a process consisting of planned activities aimed at initiating, sustaining, and realizing student development in order to facilitate learning (Açıkgöz, 2000, p.11).

As can be understood from these definitions, if we look at Figure 1, teaching is one of the subsystems or processes within the education system. One of the main reasons for confusing the concepts of education and teaching is that the ultimate goal of education, which is student development, is delivered to students during the teaching process.

Technology and Education:

In today's world, various technologies are used in every stage of education, from chalkboards to books and from projection devices to computers. As in history, the introduction of every new technology into classrooms, or in other words, into education, brings about certain challenges. Firstly, new technologies impose additional burdens on teachers.

In the Middle Ages, it was essential for teachers and students to interact face to face in education. Later, written materials emerged, and books almost replaced teachers in education. Books are still the primary educational tool today. However, books are static and insufficient materials for learners. Transferring books to electronic environments will provide easier and faster access and transform the dull face of books. Thus, the computer will enter the world of education as a new educational tool.

The increasing use of computers and the further expansion of their application areas led to the integration of computers into the education system. Initially, it was thought that computers would replace teachers and solve the problems in the education sector. However, it was soon realized that computers could not be utilized as much as expected in education. In traditional education, teachers present topics in a certain order in the classroom. The importance of presentation is the teacher's ability to convey the topics to students through speaking. During the course of instruction, supporting tools are used to deliver more topics in a short time and allow each student to personalize what is being taught. These supporting tools can be chalkboards, overhead projectors, or computers. While the teacher emphasizes important points during the presentation, they engage the students by using relevant or unrelated jokes when they sense a decline in students' interest. Interaction is present in teaching, meaning that students ask teachers questions about topics they don't understand, or teachers ask questions to students based on what they have taught.

Now let's take a look at where technology can fit into education based on what has been discussed so far. We know that presentation is the most fundamental and important element in education. If interaction and the teacher's presentation skills are crucial in the education provided, it becomes challenging to replace the teacher with a computer. However, computers can be used to support teachers' presentations. Nevertheless, numerous studies conducted comparing computer-supported education and traditional education in terms of learning outcomes have not found a significant difference between these two methods. To transition from teacher-centered education to student-centered education, utilizing computer-supported education and leveraging the internet will provide significant advantages. The internet is a place where people with similar interests come together in virtual environments, exchange information, and form new groups and communities. The internet facilitates access to people and information spread throughout the world. It can be used at all levels and in all fields of education. Conducting education over a network is easier than conducting stand-alone computer-based education because a single computer has limited information and inadequate programs, and each program has its own intricacies. On the other hand, it is easier to learn how to search for and use information in computer networks [1].

**WHAT IS DISTANCE LEARNING?**

Before defining distance learning, it would be beneficial to understand the meaning of the term "teaching." Teaching is the cornerstone of a country's technological, economic, political, and social development. As this definition implies, delivering education to the right individuals through correct methods and techniques is of vital importance for a country's future. Nowadays, the discussion has shifted from the role of education in a country's general development to how education can be delivered more effectively. Sociologists, psychologists, and experts continue to conduct research and studies on different methods and techniques in education.

Distance learning is a system that takes advantage of the benefits provided by constantly evolving technology in the field of education. The essential aspect of distance learning is that the student and teacher are located in separate geographic locations. Even if the teacher and student are under the same roof, there may still be a distance between them. This flexible definition eliminates some theoretical and stereotypical descriptions related to this program and offers a different perspective [2].

Some other definitions of distance education are as follows:

The California Distance Learning Project (CDLP) defines distance education as follows:

"A distance education program is a system that connects students with educational resources to deliver education."

The definition by the United States Distance Learning Association (USDLA) is as follows:

"The delivery of education to remote students through the use of electronic tools such as satellites, video, audio, graphics, computers, and multimedia technology." USDLA emphasizes that electronic tools or written materials and printed materials should be used in this educational program when the teacher and student are physically distant from each other [2].

As evident from these definitions, the primary goal of Internet-Based Learning (IBL) is to provide every individual with the opportunity to receive education in any field of interest without constraints of place and time. If socio-economic structure, cultural and technological development are desired in a country, priority must be given to education, and individuals must be educated to the maximum level. This is because countries advance through the people they raise, or they fall behind due to the people they fail to educate. Our goal as a country should be to catch up with the technology of developed countries. This can be achieved by providing high-level educational opportunities to anyone interested in acquiring education in their area of interest. If the goal is to ensure that every individual receives higher education, it is necessary to provide this service to individuals whether they are at home or at school. This is where the necessity and importance of distance education come into play. Distance learning or IBL provides many advantages for individuals who are obliged to work, have physical disabilities, and therefore have no chance to attend formal education but want to further develop themselves.

Determining factors of distance education:
- The physical separation of students and teachers throughout the entire education process or a significant part of it,
- The use of educational media (such as computers, phones, etc.) to facilitate communication between students and teachers and the delivery of course content to students using these tools.

### THE HISTORY OF DISTANCE LEARNING IN THE WORLD

Correspondence learning forms the basis of distance learning. Correspondence learning was used in countries like the UK, France, the US, and Germany and quickly spread. Business circles, associations, and many other organizations effectively benefited from correspondence learning. Later on, with the astonishing advancements in internet and multimedia technology, a new era began in distance education. Many universities, recognizing the importance and necessity of the internet in education, have prepared various online courses and programs.

### DISTANCE LEARNING IN TURKEY

While there have been countries that have been providing education over the internet for many years worldwide, this technology has only recently started to be used in Turkey. However, this relatively new education method in our country faces various challenges. One striking example is the inadequacy of internet infrastructure. Nevertheless, in the near future, with the increasing use of computers and the internet and the rapidly developing internet infrastructure, there will be more opportunities to work on different aspects of education. Internet-based learning provides the following benefits:

- Flexibility
- Interactivity
- Accessibility
- Reusability and shareability
- Upgradability [3].

# THE ROLE OF IBL IN TODAY'S EDUCATION SYSTEM AND TRADITIONAL EDUCATION

Interest in Internet-Based Learning (IBL) is increasing day by day. Although it receives significant attention in the academic field, there are still doubts surrounding IBL. Will IBL replace classroom-based education, or will it coexist with traditional education? Such questions continue to be subjects of debate. Ultimately, what matters is whether this education will have a positive impact on students.

Putting aside these debates, one question comes to mind: If the foundation of education, understanding, and power lies in information, then the internet and computers, which allow instant access to this information from anywhere in the world, have a crucial role in education.

In traditional education, classroom discussions are supported by body language and different tones of voice. While this may be challenging to achieve in IBL, participants can take their time to think before responding to a question and individuals who struggle with face-to-face communication can express their ideas comfortably. For example, individuals who assert themselves dominantly in a classroom discussion may not be able to do so in an online environment. In online education, students have the opportunity to provide feedback or revisit past discussions or topics. In IBL, it is important to select the right and qualified faculty members, just as in traditional universities. A professor who tries to deliver the same course in IBL as they do in traditional education will realize that they are not the same thing. In IBL, the instructor's role is not just to present information but to guide students towards knowledge. The differences between traditional and IBL classrooms can be summarized in Table 1 below.

| Traditional Education Environment | IBL Environment |
| --- | --- |
| Lesson-based | Discussion-based |
| Structural | Flexible |
| Goal-oriented | Outcome-oriented |
| Mostly teacher-centered | Independent student |
| Large classrooms | Small classrooms |
| Teacher is the source of information | Teacher guides towards knowledge |

Table 1. Comparison of traditional education and IBL.

In IBL, students have a greater responsibility towards the course compared to traditional education students. They should generate ideas and contribute to the discussion topics presented by the teacher in order to better understand the subject. This is the foundation of IBL. Students should actively participate in the class rather than being passive individuals. They analyze and learn through class discussions. Unlike in traditional education, the size of classes can also be considered in IBL. It is a well-known fact that our education system has always complained about overcrowded classrooms, and the same issues persist in traditional education today. Both the lack of teachers and inadequate classrooms in universities prevent the formation of small student groups. Even if such an opportunity exists, it places an excessive burden on faculty members. In IBL, it is easier to form and interact with small student groups. Instructors can participate in more than one discussion forum instead of just one.

## DISADVANTAGES OF IBL

One of the questions that arises about IBL is that students cannot have face-to-face communication with their peers and teachers. Another issue in IBL is related to courses where hands-on skills, such as chemistry or electronics, are crucial. The question of whether internet-based learning is sufficient for courses where practical applications play a significant role remains a separate topic of discussion.

## REASONS FOR CHOOSING IBL

Before discussing why IBL is chosen as an educational method, it is necessary to examine who makes this choice. IBL is a result of distance education. Its foundation lies in students having the opportunity to attend classes at any time and from anywhere they desire. The majority of students who choose this education method are around 26 years old and employed individuals. Their goal is to continue their education while continuing their jobs. Additionally, women constitute the majority in this education system. Women who need to spend more time at home and are unable to leave home choose IBL.

## CONCLUSION

The main theme that this article aims to convey is the importance of education in order to keep up with rapidly advancing technology and secure our position among developed countries. If we look at other developed countries around us as examples, they all have a say in the field of technology. To be influential in the field of technology, educated individuals are needed. However, in our country, due to social and economic conditions, many people are forced to quit their education at an early age to work. Perhaps every individual who is forced to quit education actually hinders our progress in the field of technology as a country. This is where IBL comes into play, providing these individuals with educational opportunities once again. It also offers alternatives to many people who would otherwise have to quit their education due to their jobs. Distance should not be an issue in IBL. It should be remembered that even if the teacher and student are miles apart geographically, they can be close to each other, while even if they are in the same classroom, they can be miles away from each other. Everything depends on the student's desire to learn.